\documentclass[aps,prb,article,twocolumn,showpacs,longbibliography,preprintnumbers,amsmath,amssymb,superscriptaddress]{revtex4-2}

\usepackage{graphicx}
\usepackage{color}
\usepackage{mathtools}
\usepackage{amsmath}
\usepackage{bbold}
\usepackage{verbatim}
\usepackage{physics}
\usepackage{bm}
\usepackage{bbm}
\usepackage{soul}

\begin{document}

\title{Neodymium ions as charge reservoir in NdNiO$_2$:\\
from lack of long range order to electron-doping-induced antiferromagnetism}

\author{Adam K\l{}osi\'nski}
\email{adam.klosinski@fuw.edu.pl}
\affiliation{
\mbox{Institute of Theoretical Physics, Faculty of Physics, 
University of Warsaw, Pasteura 5, PL-02093 Warsaw, Poland}}
\affiliation{
\mbox{International Research Centre MagTop, Institute of Physics PAS, Aleja Lotnik\'ow 32/46, PL-02668 Warsaw, Poland}}

\author{Roman Drachynskyi}
\affiliation{\mbox{Institute of Theoretical Physics, Jagiellonian University, \L{}ojasiewicza 11, PL-30348 Krak\'ow, Poland}}
\affiliation{\mbox{Doctoral School of Exact and Natural Sciences, Jagiellonian University, \L{}ojasiewicza 11, PL-30348 Krak\'ow, Poland}}

\author{Krzysztof Wohlfeld$\,$}
\affiliation{\mbox{Institute of Theoretical Physics, Faculty of Physics, University of Warsaw, Pasteura 5, PL-02093 Warsaw, Poland}}

\author{Wojciech Brzezicki}
\affiliation{\mbox{Institute of Theoretical Physics, Jagiellonian University, \L{}ojasiewicza 11, PL-30348 Krak\'ow, Poland}}

\begin{abstract}
We study magnetism in the electron-doped infinite-layer nickelate NdNiO$_2$. We perform an unrestricted Hartree-Fock calculation for a tight-binding model which contains both nickel and neodymium orbitals. We reproduce the self-doping effect, which is the escape of charge onto the neodymium bands. By fixing all free parameters to realistic values we find that undoped NdNiO$_2$ lies right {\it outside} the antiferromagnetic (AFM) region of the phase diagram. This is consistent with experiments, which find no long-range order in the ground state of NdNiO$_2$, yet see short-range AFM correlations and broad magnetic excitations. 
We also find that the self-doping effect leads to a dramatic increase in the stability of the AFM solution upon electron doping -- a behavior that is strikingly different from what is, for instance, observed in the cuprates.
Finally, for smaller charge transfer energies than suggested for NdNiO$_2$, the self-doping effect may be quite strong and stabilise various stripe configurations already on the mean-field level.
\end{abstract}

\maketitle

\section{Introduction}
\label{sec:introduction}

Since the 2019 discovery of high temperature superconductivity in the hole-doped Nd$_{1-x}$Sr$_{x}$NiO$_2$~\cite{Li2019}, inifinite-layer nickelates have become an important focal point for research on that phenomenon. The few years that passed since the initial discovery have seen numerous experimental and theoretical studies~\cite{Hirsch2019,Jiang2019,Hu2019,Hepting2020,Zeng2020,Karp2020,Katukuri2020,Xiang2021,Hsu2021,Zeng2022,Lee2023,Wang2023,Di2024} and a few review articles~\cite{Pickett2021,Nomura2022,Wang2024}. The common theme one encounters in this body of work is trying to understand superconductivity in the nickelates by identifying similarities with the more famous cuprate superconductors. This is because the most studied parent compound to the cuprates, La$_2$CuO$_4$, is isostructural to NdNiO$_2$ and its transition metal ion Cu$^{2+}$ is isovalent with the Ni$^{+}$ ion in the nickelate. 

Superconductivity in the cuprates is still a very controversial subject, and the debate about the nature of the superconducting mechanism there has by no means been resolved~\cite{Keimer2015}. The hope is that having found NdNiO$_2$ -- a similar and yet subtly different system, where unconventional superconductivity is also observed -- will perhaps help pin down the exact features of cuprates and nickelates which are necessary for high temperature superconductivity. At the same time, it might help identify the features which are largely irrelevant.

Some striking differences between the two superconductors are already apparent. So far, no antiferromagnetic (AFM) long-range order (LRO) has been observed in undoped NdNiO$_2$. Broad magnetic excitations~\cite{Lu2021} and short-range magnetic correlations~\cite{Fowlie2022}, both of the antiferromagnetic type, were observed instead. This sparked a discussion on the importance of Mott physics and magnetism in the infinite-layer nickelates in general as well as in their role in achieving suprisingly high-$T_C$ superconductivity~\cite{Fowlie2022, DiCataldo2024, Worm2024}. The latter is based on the fact that proximity to an AFM Mott insulator has long been touted as a key ingredient for high-$T_C$ superconductivity~\cite{Phillips2006,Phillips2009,Zaanen2011}. In this light, the lack of AFM LRO in the parent compound to the superconducting nickelate is a very interesting fact. 

\begin{figure}[t!]
    \includegraphics[width=\columnwidth]{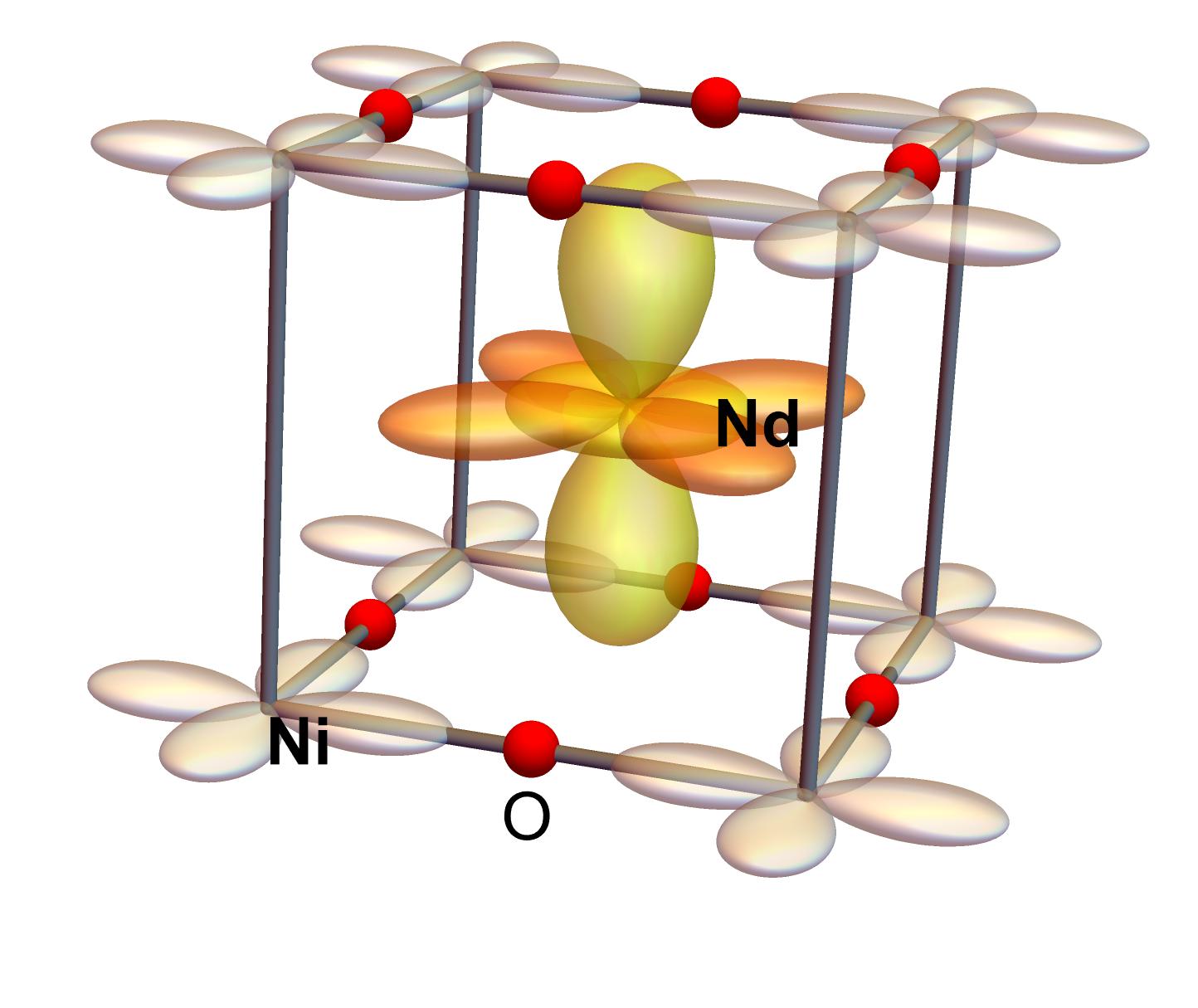}
    \caption{
    {
    Visual representation of the cubic crystal structure of NdNiO$_{2}$, showing the Ni ($d_{x^2-y^2}$) in whitish and Nd ($d_{z^2}$, $d_{xy}$) in yellow and orange colours respectively. Oxygen atoms are shown in red.}}
    \label{fig:NiNdO2}
\end{figure}

What might be the reason for lack of AFM LRO in NdNiO$_2$, a compound isostructural and isovalent to La$_2$CuO$_4$? Many have pointed to the magnitude of the Cu-O and Ni-O charge-transfer gaps, the one in nickel being significantly larger~\cite{Jiang2020}. This leads to magnetic interactions in NdNiO$_2$ being an order of magnitude weaker than in La$_2$CuO$_4$. Another possibility is the hybridisation between nickel and neodymium, much stronger than the negligible copper-lanthanum hybridisation in the cuprates~\cite{Zhang2020}. In fact, there are already several experimental studies that point to nickel-neodymium hybridisation as being significant in infinite-layer nickelates~\cite{Lu2021,Tam2022}.

In this paper we focus on the nickel-neodymium hybridisation and investigate the escape of charge from the Ni-O plane onto neodymium. For some time now, it has been a matter of contention which orbitals visible in the DFT band structure obtained for NiNdO$_2$~\cite{Lee2004,Botana2020,Nomura2022, Si2024} are necessary to properly describe its physics, in particular magnetism. Here, we will follow the article by Wu \textit{et. al.}~\cite{Wu2020}, where the authors propose a tight-binding (TB) model with three orbitals: neodymium $d_{3z^2-r^2}$ and $d_{xy}$ and nickel $d_{x^2-y2}$. This model was previously used to demonstrate a $d_{x^2-y^2}$ type pairing instability and superconducting gap in doped NiNdO$_2$, analogous to the superconductivity in copper oxides~\cite{Wu2020}. We study magnetism in the same model in the commensurate filling. We also investigate electron-doping, as in this regime the suppression of the superconducting gap was found to be much stronger than in the hole-doped regime~\cite{Wu2020}, which might suggest emerging magnetism.

\section{Model and Methods}
\label{sec:model}

We consider the tight-binding Hamiltonian
\begin{equation} \label{eq:ham-abstract}
    \mathcal{H} = \mathcal{H}_t + \mathcal{H}'.
\end{equation}
$\mathcal{H}_t$ is the kinetic Hamiltonian, which was used in \cite{Wu2020} and kindly provided by the authors of said publication. It describes the effective, long-range hopping in the infinite layer nickelate NdNiO$_2$, integrating out the oxygen orbitals and mapping the band structure onto two neodymium orbitals, $d_{3z^2-r^2}$ and $d_{xy}$ and a single nickel orbital, $d_{x^2-y^2}$. While the full tight-binding model used in \cite{Wu2020} is defined on a three-dimensional lattice, we have modified it by truncating the system to a single nickel layer and a single neodymium layer.
\begin{equation}
    \mathcal{H}_t = \vec{c}^\dag H_t \vec{c},
\end{equation}
where $\vec{c}$ is the vector of annihilation operators, whose components span all annihilation operators $c_{i \alpha \sigma}$, where $i$ numbers the unit cells, $\alpha$ is the orbital index ($1\!=\!d_{3z^2-r^2}$, $2\!=\!d_{xy}$, $3\!=\!d_{x^2-y^2}$) and $\sigma$ is the spin index. Meanwhile,

{
\begin{equation} \label{eq:ht}
    H_t = \sum_{i_x,i_y}\sum_{i_x^{\prime},i_y^{\prime}} |i_x,i_y\rangle\langle i_x^{\prime},i_y^{\prime}|  \otimes h_{i_x-i_x^{\prime},i_y-i_y^{\prime}} \otimes \sigma_0
\end{equation}
describes hopping between unit cells at positions $(i_x,i_y)$ and $(i_x^{\prime},i_y^{\prime})$.
Matrices $h$, each of
dimension three, describe hopping elements in the orbital sector and $\sigma_0$, identity in the spin sector.
In reciprocal space this becomes
\begin{equation} \label{eq:ht-q-space}
    H_t = \sum_{\mathbf{k}} \sum_{\mathbf{d}} e^{i \mathbf{k}\cdot\mathbf{d}} |k_x,k_y\rangle\langle k_x,k_y| \otimes h_{d_x,d_y} \otimes \sigma_0,
\end{equation}}
The kinetic energy is quite rich, for it follows a realistic tight-binding model with hopping elements up to third neighbor~\cite{Wu2020} and therefore we need nine matrices $h$ to describe it. We provide those in Appendix~\ref{sec:appendix1}, Eq.~\eqref{eq:matrices}. Entries of all $h$'s  are given in eV and follow \cite{Wu2020}; matrices not given in \eqref{eq:matrices} are zero.

$\mathcal{H}'$ is the Coulomb interaction on nickel atoms. In second quantization it is
\begin{equation} \label{eq:hu}
    \mathcal{H}' = U \sum_{i_x,i_y} n_{(i_x,i_y) 3 \uparrow} n_{(i_x,i_y) 3 \downarrow},
\end{equation}
with 
\begin{equation}
   n_{(i_x,i_y)3\sigma} = c^\dagger_{(i_x,i_y) 3 \sigma} c_{(i_x,i_y) 3 \sigma}. 
\end{equation}
Here operator $c^\dagger_{(i_x,i_y) 3 \sigma}$ creates an electrons in the unit cell at point $(i_x,i_y)$ in orbital labeled as $3$ (or $d_{x^2-y^2}$) with spin $\sigma$.
We treat $\mathcal{H}'$ \eqref{eq:hu} in two ways. We use  unrestricted Hartree-Fock (UHF) and restricted Hartree-Fock (RHF). We assume $U=4.5$ eV.
Note that this value gives
the $U/t \approx 12 $ (where $t$ is the Ni-Ni nearest neighbor hopping element), basically in agreement with the recently postulated ratio 
used to describe the RIXS experiments on nickelates~\cite{Rosa2024}.

While we fix $U$ as well as all hopping and on-site energy parameters to the values mentioned above, we introduce three tunable parameters in the model. This should allow us to better understand
the physics behind the model. 
The first is the total filling $n$. The model \eqref{eq:ham-abstract} contains three orbitals and the total number of six states in the unit cell. The physical filling is one electron per unit cell~\cite{Nomura2022}. (Note that the other eight electrons on Ni$^+$ ion occupy the lower lying $3d$ orbitals that are not considered here.) In our study we vary $n$ between the physical filling of $n = 1$ and the maximal value of $n = 2$, or two electrons per unit cell.
The second is the charge-transfer energy between nickel and neodymium atoms, $\epsilon$ [see Eq. \eqref{eq:matrices} in Appendix \ref{sec:appendix1}]. The range for $\epsilon$ will be given later, as it varies for different calculations.
Third, we also introduce parameter $\gamma$ which scales the nickel-neodymium hopping ($\gamma=1$ restores the realistic values suggested in~\cite{Wu2020}).  

\subsection{Unrestricted Hartree-Fock approximation} \label{subsec:uhf}

We treat the interaction term using mean-field decoupling. The mean-field decoupled term in the UHF approach reads
\begin{equation}
    \mathcal{H'} \approx \mathcal{H}'_{UHF} + \Delta E_{UHF},
\end{equation}
\begin{equation} \label{eq:uhf}
    \mathcal{H}'_{UHF} = \vec{c}^\dag H'_{UHF} \vec{c},
\end{equation}
with
{
\begin{align} \label{eq:uhf-1-particle}
    \begin{split}
        H'_{UHF} &= U \sum_{i_x,i_y} |i_x,i_y\rangle\langle i_x,i_y| \otimes |3\rangle\langle 3|\\
        &\otimes \Big[ \frac{1}{2} \rho_{(i_x,i_y)\downarrow} \left( \sigma_0 + \sigma_z \right) + \frac{1}{2} \rho_{(i_x,i_y)\uparrow} \left( \sigma_0 - \sigma_z \right)\\
        &- \Re[ \Delta_{(i_x,i_y)}] \sigma_x - \Im[ \Delta_{(i_x,i_y)} ] \sigma_y \Big].
    \end{split}
\end{align}
In the above, the first term of the product is the projector onto the unit cell at $(i_x,i_y)$, the next one is the projector onto the nickel $d_{x^2-y^2}$ orbital and $\sigma_\alpha, \; \alpha=x,y,z$ are the standard Pauli matrices. Furthermore,
\begin{equation}
    \begin{array}{c}
    \rho_{(i_x,i_y)\sigma} = \langle n_{(i_x,i_y)\sigma} \rangle\\
    \Delta_{(i_x,i_y)} = \langle c^\dagger_{(i_x,i_y)\uparrow} c_{(i_x,i_y)\downarrow} \rangle,
    \end{array}
\end{equation}
where $\langle \cdot \rangle$ indicates the ground-state expectation value. The constant energy shift is
\begin{equation}
    \Delta E_{UHF} = U \sum_{i_x,i_y} \left( -\rho_{(i_x,i_y)\uparrow} \rho_{(i_x,i_y)\downarrow} + |\Delta_{(i_x,i_y)}|^2 \right). 
\end{equation}
}

\subsection{Restricted Hartree-Fock approximation.} \label{subsec:rhf}

In the RHF approach, we assume $\Delta_{(i_x,i_y)} = 0$. We plug this into \eqref{eq:uhf} and go to reciprocal space
{
\begin{equation}
    \begin{array}{c}
    \rho_{\mathbf{k}\sigma} =\frac{1}{N} \sum_{\mathbf{r}} e^{i \mathbf{k} \cdot \mathbf{r}} \rho_{\mathbf{r}\sigma}\\
    c_{\mathbf{k}\sigma} = \frac{1}{\sqrt{N}} \sum_\mathbf{r} e^{i \mathbf{k} \cdot \mathbf{r}} c_{\mathbf{r}\sigma},
    \end{array}
\end{equation}}
Furthermore, we put restriction on $\rho_{\mathbf{k}\sigma}$, limiting ourselves to ferromagnetic (FM) and antiferromagnetic (AFM) instabilities, namely
\begin{equation}
    \begin{array}{c}
    \rho_{\mathbf{k}\uparrow} = \rho_{0\uparrow} \delta(\mathbf{k}) + \rho_{1} \delta(\mathbf{k} - (\pi,\pi))\\
    \rho_{\mathbf{k}\downarrow} = \rho_{0\downarrow} \delta(\mathbf{k}) - \rho_{1} \delta(\mathbf{k} - (\pi,\pi)).
    \end{array}
\end{equation}
From this, we obtain
\begin{equation} \label{eq:rhf}
\mathcal{H}' \approx \mathcal{H}'_{RHF} + \Delta E_{RHF},
\end{equation}
\begin{equation}
\mathcal{H}'_{RHF} = \vec{c}^\dag H'_{RHF} \vec{c},
\end{equation}
with
{
\begin{align} \label{eq:rhf-1-particle}
    \begin{split}
    H'_{RHF} &= U \sum_\mathbf{k} \Big\{ |k_x,k_y\rangle\langle k_x,k_y| \otimes |3\rangle\langle 3|\\
    &\otimes \left[ \frac{1}{2} \rho_{0\downarrow} \left( \sigma_0 + \sigma_z \right) + \frac{1}{2} \rho_{0\uparrow} \left( \sigma_0 - \sigma_z \right) \right]\\
    &+ |k_x,k_y\rangle\langle k_x\!+\!\pi,k_y\!+\!\pi| \otimes |3\rangle\langle 3| \otimes \rho_1 \sigma_z \Big\}.
    \end{split}
\end{align}}
The first term in \eqref{eq:rhf-1-particle} corresponds to the FM instability (Stoner mechanism~\cite{Stoner1939}), while the second term describes the AFM instability. The constant energy shift in the RHF approach is
{
\begin{align} \label{eq:constant-rhf}
\begin{split}
    \Delta E_{RHF} &= - U \sum_\mathbf{r} \rho_{\mathbf{r}\uparrow} \rho_{\mathbf{r}\downarrow} = - U \sum_\mathbf{k} \rho_{\mathbf{k}\uparrow} \rho_{\mathbf{k}\downarrow}\\
    &= U N \left( - \rho_{0\uparrow} \rho_{0\downarrow} + \rho_1^2 \right).
\end{split}    
\end{align}}

\subsection{The modified Broyden Mixing}
\label{subsec:broy}

Solving self-consistent equations with many mean fields, which is the case for UHF approach, is a computationally demanding and often slow process. To deal with complications, Broyden's method was modified to make better convergence properties \cite{Zitko09}. This method uses $M$ previous iterations of the self-consistency loop to update the approximate inverse Jacobian. The full expression for this method is the following:
\begin{equation}
    \boldsymbol{V}^{m+1} = \boldsymbol{V}^{m} + x \boldsymbol{F}^{m} - \sum_{n=1}^{m-1} \sum_{k=1}^{m-1} \left( \omega_{n} \omega_{k} c_{k}^{m} \beta_{k,n}^{m} \boldsymbol{U}^{n} \right). 
    \label{eq:broy}
\end{equation}
Here $\boldsymbol{V}^{m}$ is a vector of input mean fields, i.e. the mean fields that are fed into the Hamiltonian, at $m$-th iteration and $\boldsymbol{F}^{m}$, being the residual vector, is the difference between the recalculated mean fields, after the Hamiltonian is diagonalized, and the input mean fields $\boldsymbol{V}^{m}$. Reaching self-consistency means that $\boldsymbol{F}^{m}=0$. The first part of Eq. (\ref{eq:broy}) represents simple linear update where $x$ is a mixing parameter. The next part is the Broyden's mixing correction, which ensure that the calculations are going in the right direction in the phase space. The matrix elements:
\begin{equation}
    {c}^{m}_{k} = (\Delta\boldsymbol{F}^{k})^{\dagger} \boldsymbol{F}^{m},
\end{equation}
quantify the overlap between the current residual vector and the previous one and
\begin{equation}
     \boldsymbol{U}^{n} = x\Delta\boldsymbol{F}^{n} + \Delta\boldsymbol{V}^{m},
\end{equation}
is a secant update vector. We also need:
\begin{equation}
    \beta_{k,n}^{m} = [(w_{0}^{2} \boldsymbol{1} + A^{m})^{-1}]_{k,n},
\end{equation}
and
\begin{equation}
A^{m}_{k,n}=w_kw_n(\Delta\boldsymbol{F}^{n})^{\dagger}\Delta\boldsymbol{F}^{k}.
\end{equation}
The matrix $A^{m}_{k,n}$ measures pairwise overlaps between the normalized residual vectors, while $\beta_{k,n}^{m}$ provides the corresponding weights for the update direction. The regularization term $w_{0}^{2} \boldsymbol{1}$ ensures numerical stability by preventing singularities in the matrix inverse operation. The final definitions are:
\begin{equation}
    \Delta\boldsymbol{F}^{n} = \frac{\boldsymbol{F}^{n+1}-\boldsymbol{F}^{n}}{|\boldsymbol{F}^{n+1}-\boldsymbol{F}^{n}|},
\end{equation}
\begin{equation}
    \Delta\boldsymbol{V}^{m} = \frac{\boldsymbol{V}^{m+1}-\boldsymbol{V}^{m}}{|\boldsymbol{F}^{m+1}-\boldsymbol{F}^{m}|}.
\end{equation}
Symbol $\boldsymbol{1}$ in equation for $\beta^m$ denotes $(m-1)\times(m-1)$ identity matrix. The weights $w_{k}, w_{n}$ are usually equal to 1. The regularization parameter  $w_{0}$ is usually fixed to 0.01. In our implementation it is adaptively updated at each iteration according to the condition number of $A^{m}_{k,n}$ in order to avoid instabilities and improve convergence robustness.

Note that the correction part of the Broyden mixing can use a finite number of previous iterations. We choose this number to be 100 which means that after 100 iterations the history is reset and rebuilt. This choice reduces sensitivity to the random initialization, keeps the cost of calculation controlled, and prevents the mixer from overfitting stale history -- issues that can stall convergence to an undesired fixed point. Restarting the history can also help the algorithm escape metastable solutions.

\section{Results} 
\label{sec:results}

\subsection{Restricted Hartree-Fock} \label{subsec:rhf-res}

\begin{figure}[t!]
    \includegraphics[width=\columnwidth]{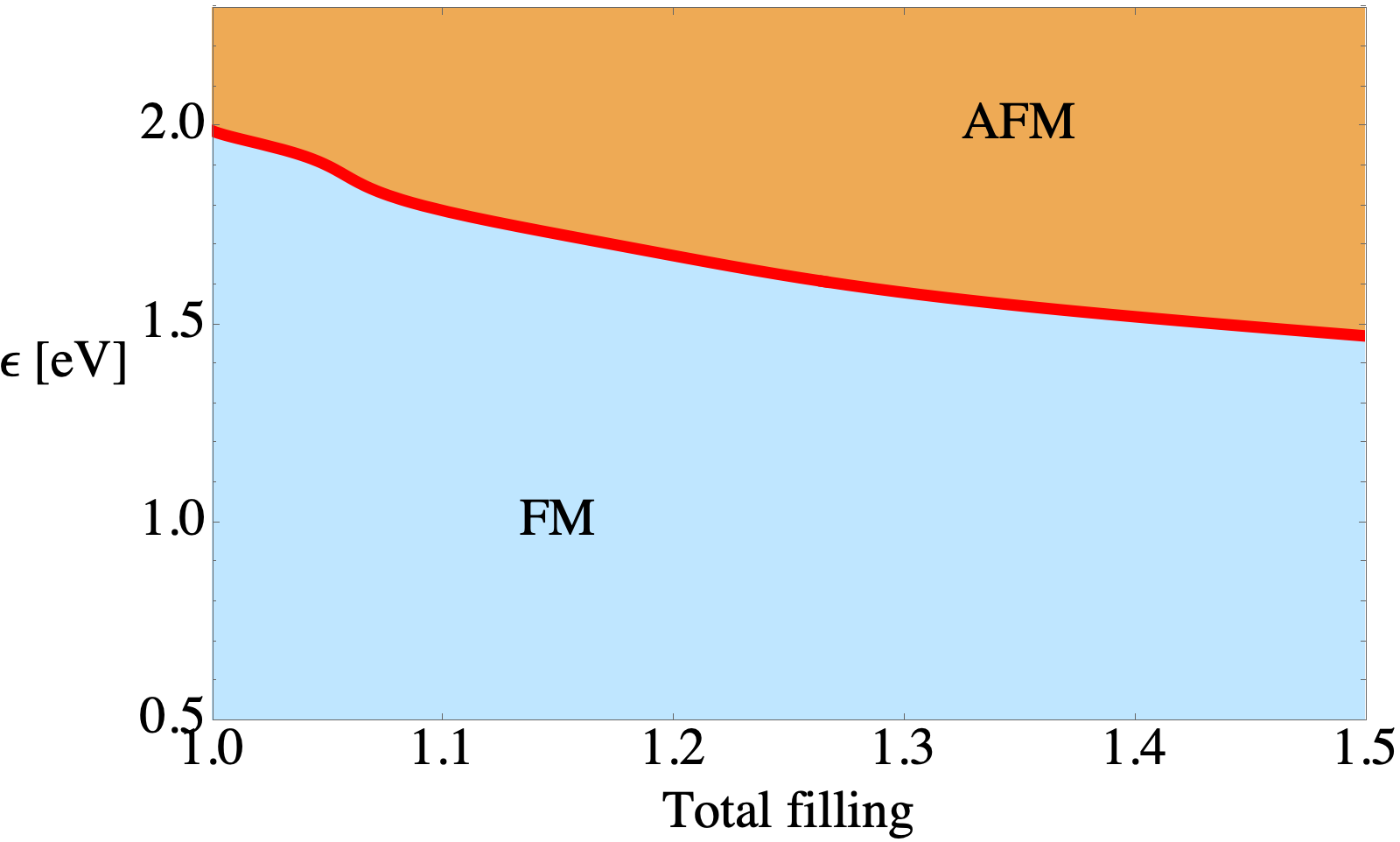}
    \caption{The phase diagram of  model \eqref{eq:ham-abstract} in the RHF approximation, that is with $H_t$ given by \eqref{eq:ht-q-space} and $H'$ given by $H'_{RHF}$ \eqref{eq:rhf-1-particle}. The red line separates the FM and the AFM phases. On that line, the energies of the FM and the AFM phases are equal. The calculation was done for a system on a pseudomomentum grid of size $10\times10$. The parameters $(\epsilon, n)$ were taken from a grid with steps $\delta\epsilon = 0.03$ eV and $\delta n = 1/600 \approx 0.0017$.}
    \label{fig:restricted-pd}
\end{figure}

In order to establish the energy landscape of the FM and AFM phases in the tight-binding model of the infinite-layer nickelate NdNiO$_2$ described above, we applied the RHF method. Namely, we solved the model \eqref{eq:ham-abstract} with $H_t$ given by \eqref{eq:ht-q-space} and $H'$ given by $H'_{RHF}$ \eqref{eq:rhf-1-particle}. There were three self-consistently determined parameters, $\rho_{0\uparrow}$ $\rho_{0\downarrow}$ and $\rho_1$, determined using the standard iterative procedure with linear mixing
\begin{align} \label{eq:iterative-procedure}
    \begin{split}
        \rho_{0\sigma}^l &= \left(1-x\right) \rho_{0 \sigma}^{l-1} + x \frac{1}{N} \sum_{\mathbf{r}} \langle n_{\mathbf{r},\sigma} \rangle_{l-1}\\
        \rho_{1}^l &= \left(1-x\right) \rho_1^{l-1} + x \frac{1}{N} \sum_{\mathbf{r}} e^{i(\pi,\pi) \cdot \mathbf{r}} \left\langle n_{\mathbf{r},\uparrow} \right\rangle_{l-1}.
    \end{split}
\end{align}
Above, $\rho_{0 \sigma}^l$ and $\rho_1^l$ denote the values of the self-consistently determined parameters obtained after $l$ iterations, while $\left\langle \bullet \right\rangle_{l}$ denotes the expectation value taken for the ground state obtained in the $l$-th iteration. $x$ is the update parameter. We use $x=0.2$. Furthermore, for the FM phase we restricted the procedure by demanding that 
\begin{equation}
    \rho_1 = 0
\end{equation} 
in \eqref{eq:rhf-1-particle}-\eqref{eq:iterative-procedure}. Similarly, for the AFM phase we demanded that 
\begin{equation}
    \rho_{0\uparrow} = \rho_{0\downarrow} \equiv \rho_0.
\end{equation} 

The model was defined on a $10\times10$ pseudomomentum lattice. In order to construct the phase diagram, we performed a scan over two free parameters. The first free parameter was the total filling $n$. 
The second parameter was the nickel-neodymium charge transfer energy $\epsilon$ [see Eq. \eqref{eq:matrices} in Appendix \ref{sec:appendix1}]. 
We investigated the parameter range
\begin{equation} \label{eq:parameter-range-rhf}
\begin{array}{c}
\epsilon \in [1.2 \; \text{eV}, 2.3 \; \text{eV}],\\
n \in [1, 1.5],
\end{array}
\end{equation}
 with steps $\delta \epsilon = 0.03$ eV and $ \delta n = 0.01$. This range was chosen because it covers the physical point ($\epsilon = 2$ eV, $n=1$)~\cite{Wu2020}, as well as electron doped systems. It also covers a reasonably wide range of values for the $\epsilon$ parameter, whose value is uncertain. At each point $(\epsilon, n)$ within the range \eqref{eq:parameter-range-rhf} we used the iterative procedure to self-consistently determine the parameters $\rho_{0\uparrow}$, $\rho_{0\downarrow}$ (FM) or $\rho_{0}$, $\rho_1$ (AFM). The calculation was considered to converge if 
\begin{equation}
    \max \left(\rho_{0\uparrow}^{l+1} - \rho_{0\uparrow}^l, \rho_{0\downarrow}^{l+1} - \rho_{0\downarrow}^l \right) < 10^{-6}
\end{equation}
in the FM case and
\begin{equation}
    \max \left(\rho_{0}^{l+1} - \rho_{0}^l, \rho_1^{l+1} - \rho_1^{l} \right) < 10^{-6}
\end{equation}
in the AFM case. 

We found that the calculation converges in the entire free parameter range, for both FM and AFM phases, so that at each point $(\epsilon, n)$ in the free parameter space stable FM and AFM solutions were found.  At each point the self-consistently determined parameters -- either $\rho_{0\uparrow}$ and $\rho_{0\downarrow}$ (FM) or $\rho_0$ and $\rho_1$ (AFM) -- were then used to calculate the ground state energy for each of the phases. This allowed us to find a line on the $(\epsilon, n)$ plane which separates the regions where the FM and AFM solutions are the respective ground states (see Fig. \ref{fig:restricted-pd}). This line corresponds to such parameters $\epsilon$ and $n$ for which the FM and AFM solutions have equal energies.

\subsection{Unrestricted Hartree-Fock} \label{subsec:uhf-res}

\begin{figure*}[t!]
    \includegraphics[width=\columnwidth]{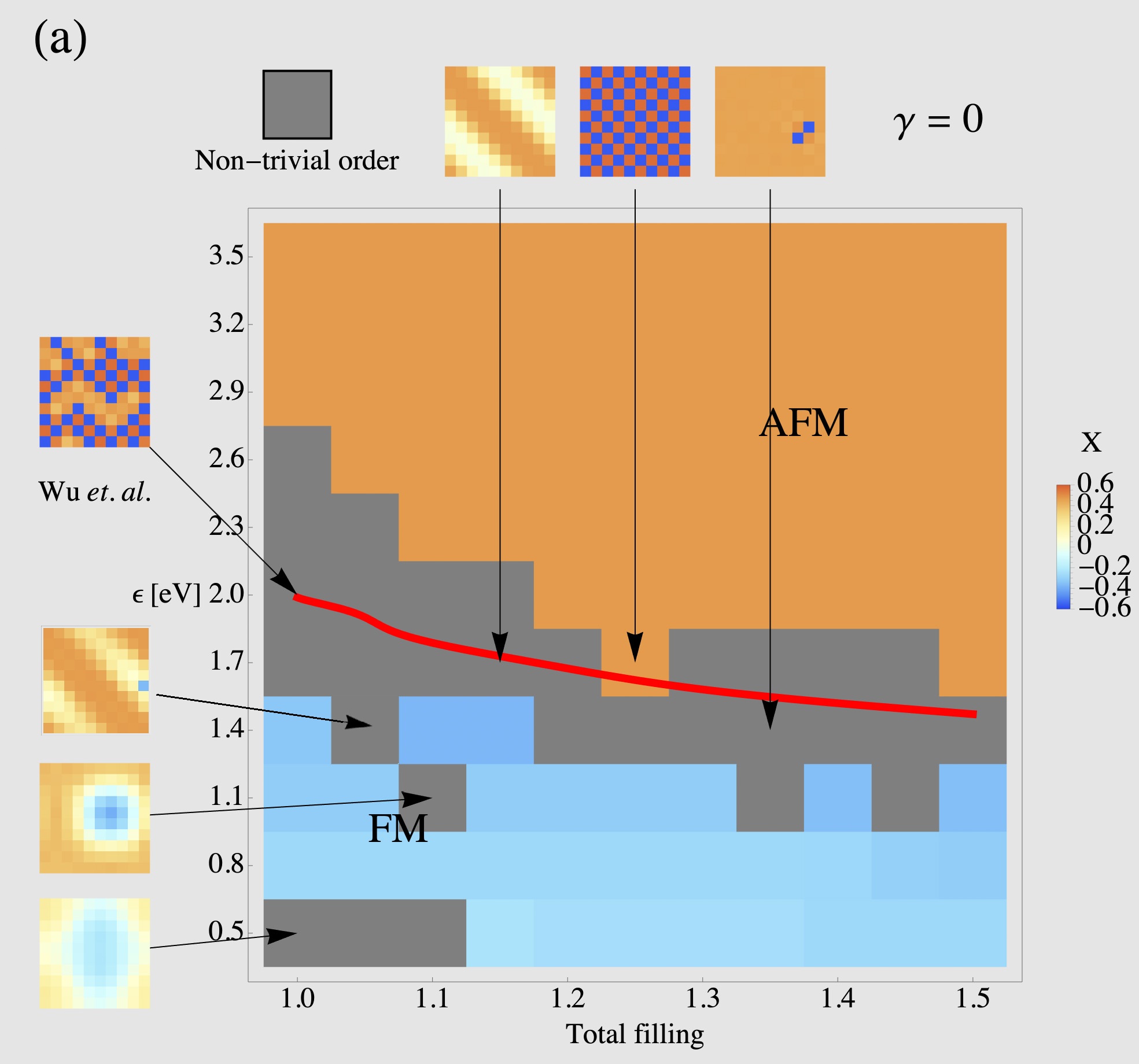}%
    \includegraphics[width=\columnwidth]{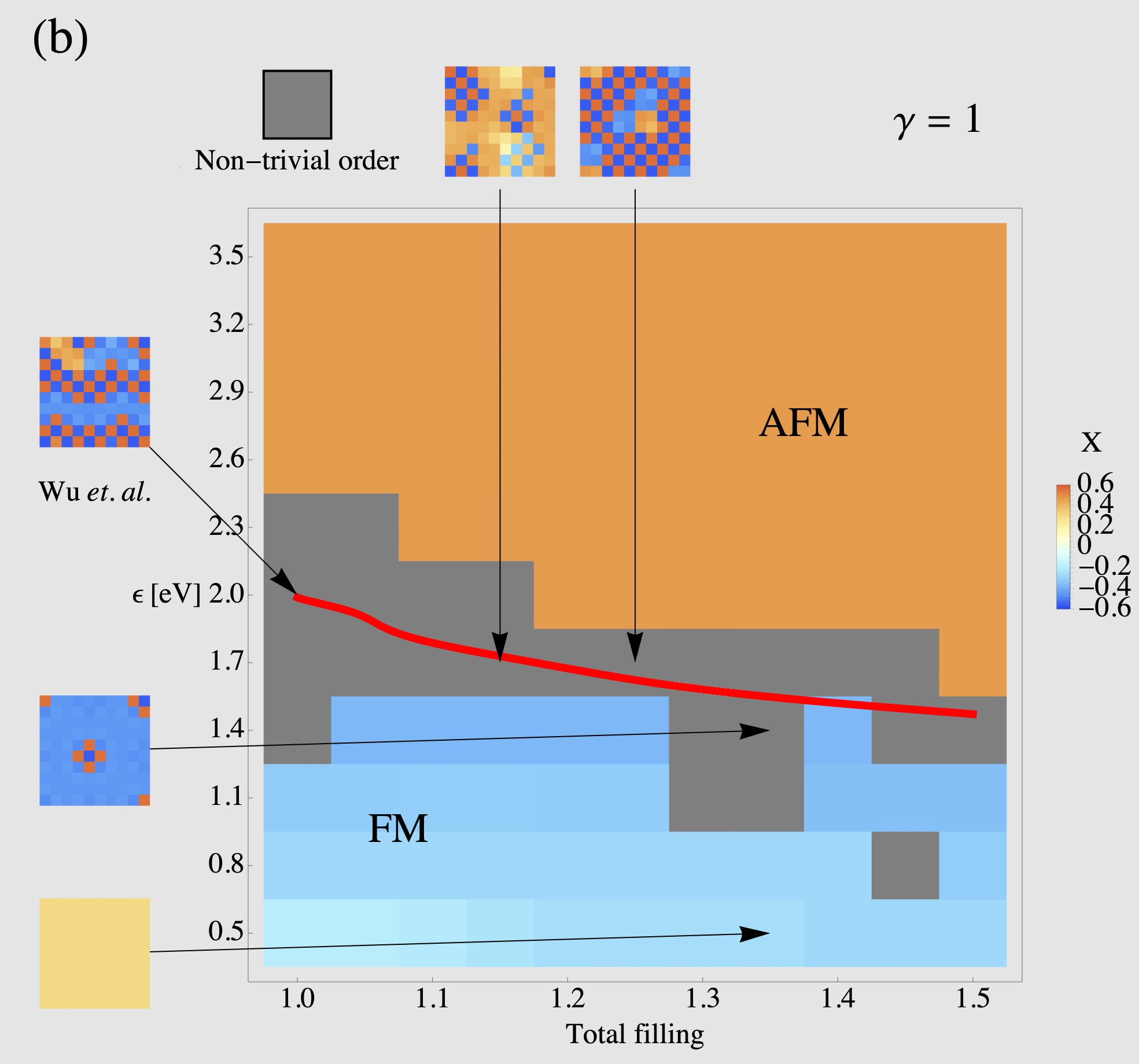}\\
    \vskip -0.5mm
    \includegraphics[width=\columnwidth]{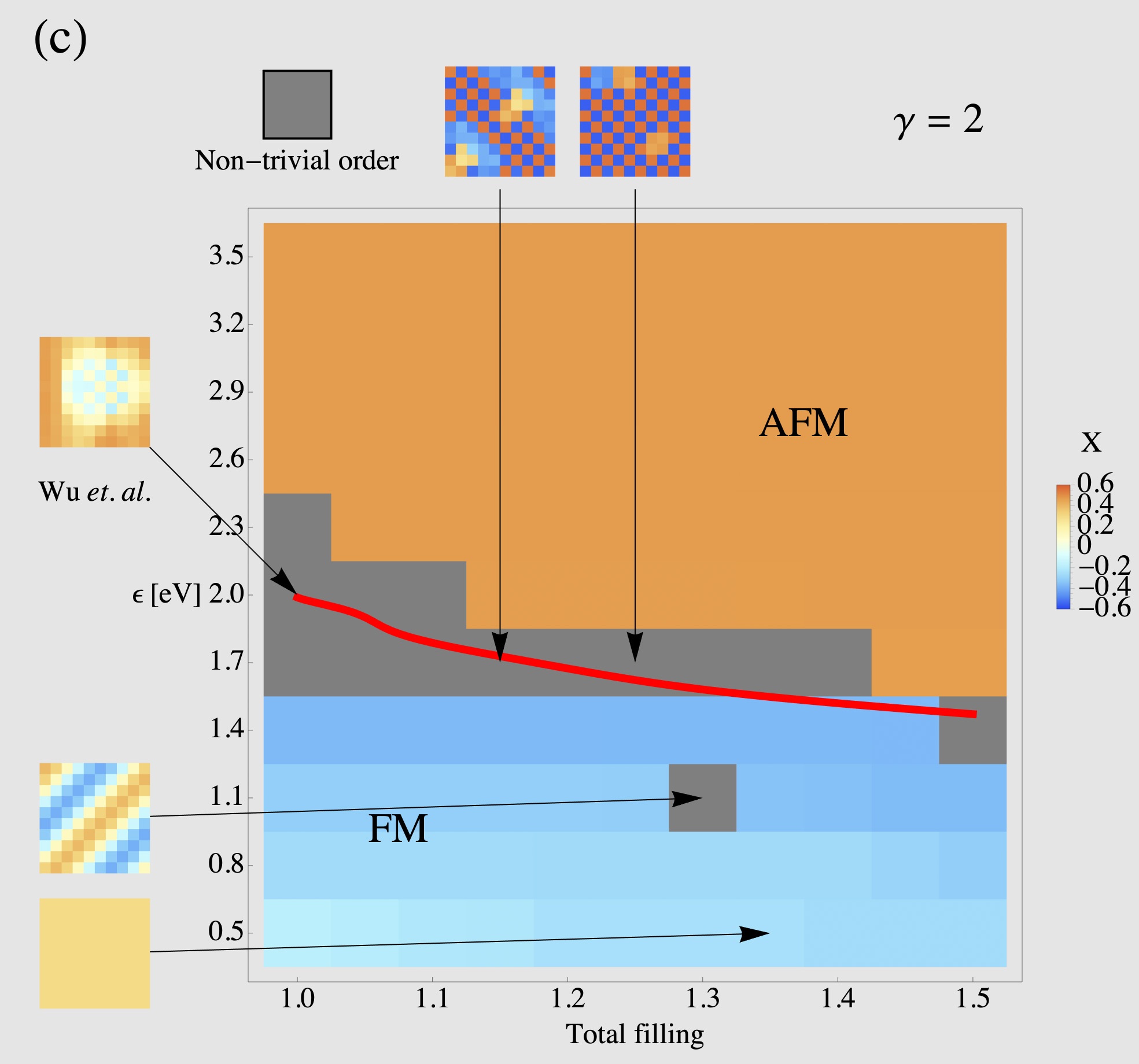}%
    \includegraphics[width=\columnwidth]{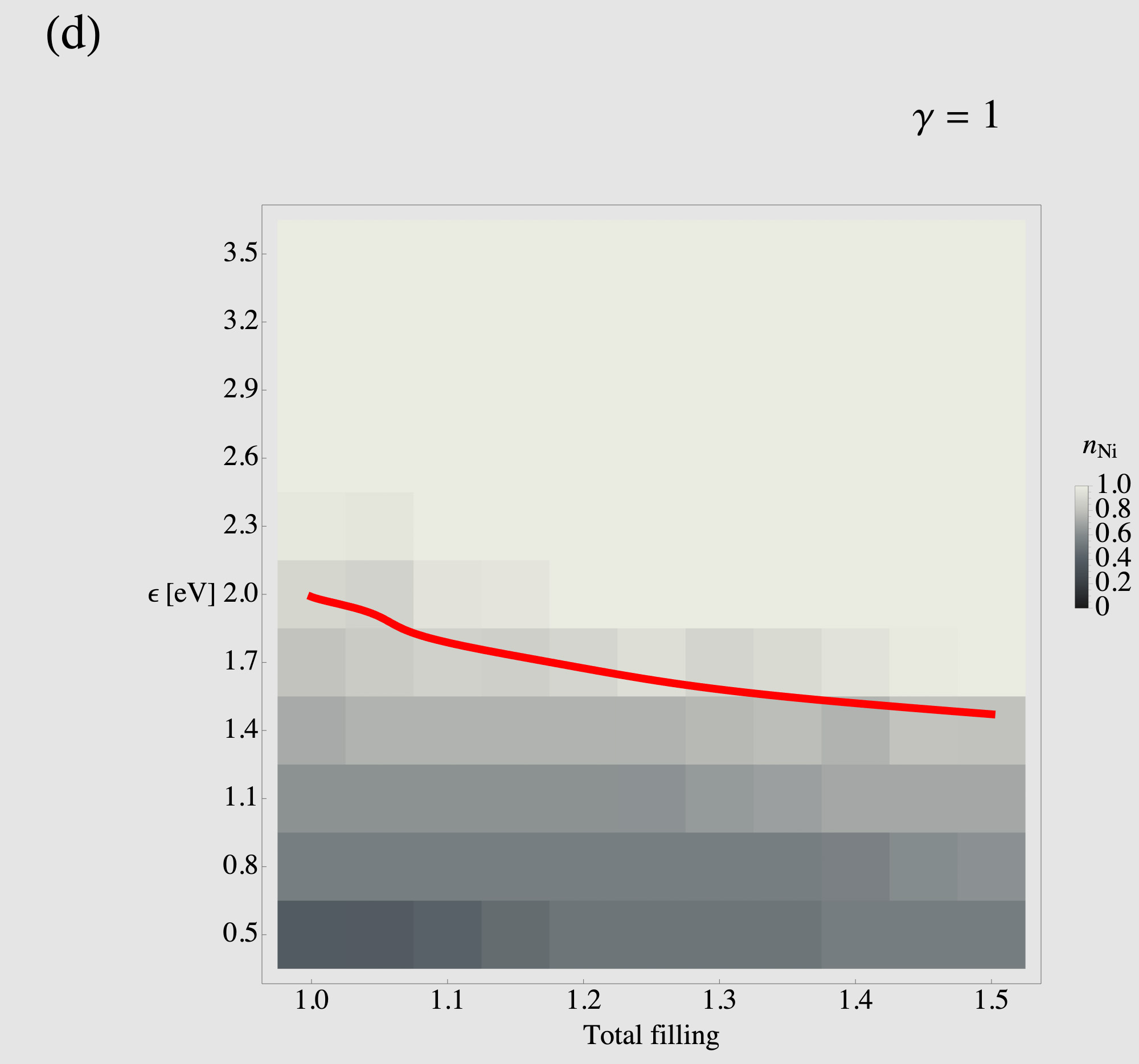}
    \caption{\textbf{Panels (a)-(c):} The value of the order parameter $X$ (see text) as a function of nickel-neodymium charge-transfer energy $\epsilon$ and total filling $n$. Results for the model \eqref{eq:ham-abstract} with kinetic energy $H_t$ given by \eqref{eq:ht} and Hubbard $H'$ given by $H'_{UHF}$ \eqref{eq:uhf-1-particle}. Results are shown for three different values of the nickel-neodymium hopping scaling parameter: (a) $\gamma = 0$, (b) $\gamma = 1$, (c) $\gamma = 2$. The insets show real-space maps of $S_z$ corresponding to particular values of $X$. The inset marked `Wu \textit{et. al.}' corresponds to $\epsilon$ and $n$ values used in~\cite{Wu2020}. The red line is the phase boundary, established using the RHF calculation (see text). The gray boxes indicate exotic (neither FM, nor AFM, nor paramagnetic) order. \textbf{Panel (d):} The filling on the nickel layer as a function of nickel-neodymium charge-transfer energy $\epsilon$ and total filling $n$, for $\gamma = 1$. The red line visible in the plot is the same line as in panels (a)-(c).}
    \label{fig:unrestricted-pd}
\end{figure*}

The UHF calculations were performed as follows. We used the Hamiltonian \eqref{eq:ham-abstract} with $H_t$ given by \eqref{eq:ht} and $H'$ given by $H'_{UHF}$ \eqref{eq:uhf-1-particle} and found its ground state using the standard iterative procedure. 
The self-consistently determined parameters were $\rho_{\mathbf{r}\uparrow}$, $\rho_{\mathbf{r}\downarrow}$ and $\Delta_\mathbf{r}$, with $\mathbf{r}=(i_x,i_y)$ being the unit cell index. The iterative step was
\begin{align} \label{eq:iterative-procedure}
    \begin{split}
        \rho_{\mathbf{r}\sigma}^l &= \left(1-x\right) \rho_{\mathbf{r} \sigma}^{l-1} + x \langle n_{\mathbf{r},\sigma} \rangle_{l-1}\\
        \Delta_{\mathbf{r}}^l &= \left(1-x\right) \Delta_{\mathbf{r}}^{l-1} + x \left\langle c^\dag_{\mathbf{r}\uparrow} c_{\mathbf{r}\downarrow} \right\rangle_{l-1},
    \end{split}
\end{align}
with linear mixing parameter $x=0.2$. 

In our calculations, we used a 10x10 lattice with periodic boundary conditions (PBC), which resulted in 300 self-consistently determined parameters at each point $(\epsilon, n)$ in the free parameter space. Because of this increased complexity with respect to the RHF calculation, we changed the steps sizes in the scan over the free parameters $\epsilon$ and $n$ to $\delta n = 0.05$ and $\delta \epsilon = 0.3$ eV. We also investigated a somewhat wider region in the free parameter space 
\begin{equation} \label{eq:parameter-range-uhf}
\begin{array}{c}
\epsilon \in [0.5 \; \text{eV}, 3.5 \; \text{eV}],\\
n \in [1, 1.5].
\end{array}
\end{equation}
During the calculation, we set an upper limit on the number of iterations, the same for each point $(\epsilon, n)$ and equal to 150000. The desired accuracy was
\begin{equation}
    \max_{\mathbf{r}\sigma} \left( \rho_{\mathbf{r}\sigma}^{l+1} - \rho_{\mathbf{r}\sigma}^l, \Delta_{\mathbf{r}}^{l+1} - \Delta_{\mathbf{r}}^{l} \right) < 10^{-6},
\end{equation}
where $\rho_{j\sigma}^l$ and $\Delta_i^l$ indicate the respective parameter values at the $l$-th step of the iterative procedure.
If for a given point in the phase diagram the calculation did not converge with prescribed accuracy within iterations limit, we used modified Broyden mixing for this point.

For each point $(\epsilon, n)$ we established the values of the 300 self-consistently determined parameters $\rho_{j\uparrow}$, $\rho_{j\downarrow}$ and $\Delta_i$, as well as the ground state. We then characterized this ground state by an order parameter $X$, defined as \begin{equation} \label{eq:order-parameter}
    X = f_{\rm AFM} - f_0,
\end{equation}
where $f_0$ is the zeroth Fourier coefficient of the discrete Fourier transform of the distribution of the observable 
$S_{\mathbf{r}z}$ on the $10\times10$ lattice,
\begin{align} \label{eq:fourier-expansion}
\begin{split}
    S_{\mathbf{k}z} &= \frac{1}{N} \sum_j e^{i \mathbf{k} \cdot \mathbf{r}} S_{jz}\\
    &= f_0 \delta (\mathbf{k}) + f_{\rm AFM} \delta \left( \mathbf{k} - (\pi,\pi) \right) + \cdots,
\end{split}
\end{align}
with
\begin{equation}
    S_{\mathbf{r}z} = \frac{1}{2} \left( \rho_{\mathbf{r}\uparrow} - \rho_{\mathbf{r}\downarrow} \right),
\end{equation}
and $f_{\rm AFM}$ is the Fourier coefficient corresponding to AFM order. Moreover, for each $(\epsilon, n)$ the higher-order Fourier coefficients in the Fourier expansion of $S_{jz}$ \eqref{eq:fourier-expansion} were also checked for non-trivial order. 

The results are shown in Fig. \ref{fig:unrestricted-pd}. In panel (a) we show the results for the model \eqref{eq:ham-abstract} with $H_t$ given by \eqref{eq:ht} and $H'$ given by $H'_{UHF}$ \eqref{eq:uhf-1-particle} and the nickel-neodymium hopping scaling parameter $\gamma = 0$. In panels (b) and (c) we show the same model, albeit with $\gamma = 1$ and $\gamma = 2$ respectively. The red line on the plot corresponds to the line in Fig.~\ref{fig:restricted-pd}. 

We see that the regions of the parameter space corresponding to the FM and AFM phases are largely the same in UHF and RHF calculations. They do not change with $\gamma$ either. Even in panel (a), where all terms coupling the nickel and neodymium atoms were removed, we see that the phase diagram remains largely the same. This corresponds to a model with two uncoupled planes -- one consisting of nickel and one of neodymium. 

Crucially, the free parameter values of $\epsilon \approx 2$ eV and $n = 1$, which are realistic for an undoped system~\cite{Wu2020}, lie exactly at the phase boundary. This suggests that magnetism in NdNiO$_2$ may be suppressed due to the competition between AFM and FM orders. 

In the vicinity of the phase boundary, already present in the RHF calculations, non-trivial phases stabilise in all panels (a)-(c), that is for all values of $\gamma$. Looking at Fig.~\ref{fig:unrestricted-pd} (a) vs (b) vs (c), one sees that the region of phase coexistence is narrowed with increasing $\gamma$, progressively replaced by the FM region. This suggests that the inter-layer coupling tends to stabilise the FM phase.

Finally, in panel (d) we show the filling on the nickel plane $n_{\rm Ni}$ as a function of $n$ and $\epsilon$, for $\gamma = 1$. This demonstrates the self-doping effect, namely the escape of electrons onto the neodymium plane. It is evident that the AFM phase perfectly correlates with the region in which $n_{\rm Ni}=1$, namely half-filling.

\subsection{Unrestricted Hartree-Fock with the modified Broyden mixing}\label{subsec:uhf-broy}

\begin{figure}[t!]
    \includegraphics[width=\columnwidth]{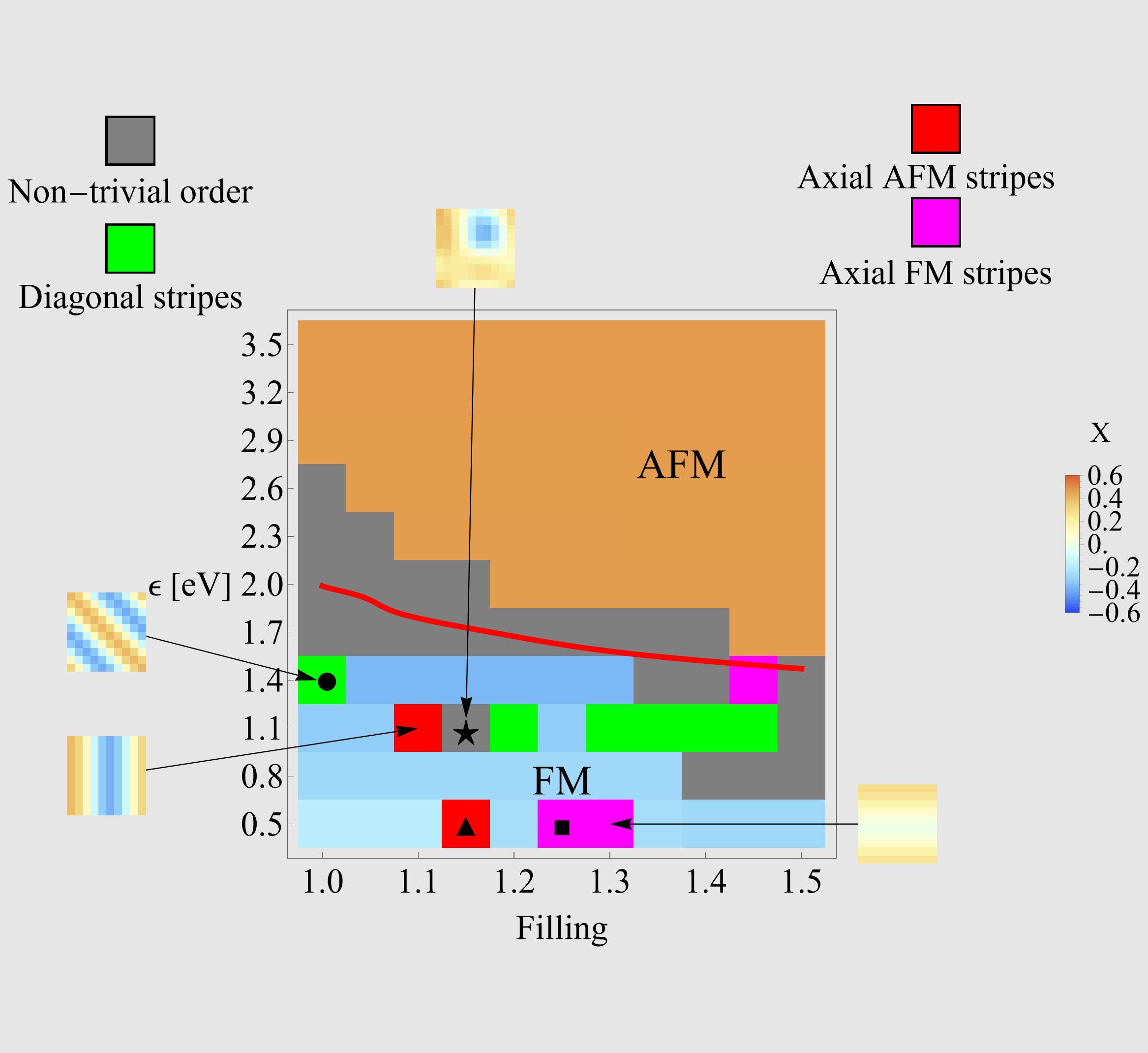}
    \caption{{
    The value of the order parameter $X$ (see text) as a function of nickel-neodymium charge-transfer energy $\epsilon$ and total filling $n$ calculated for the nickel-neodymium scaling parameter $\gamma = 1$ using the UHF approximation with the modified Broyden mixing. The red line is the phase boundary, established using the RHF calculation (see text).}}
    \label{fig:UHF-BRi}
\end{figure}

As the final step, 
we replace the linear mixing by the modified Broyden mixing in the UHF scheme. Using this method 
we perform calculations for the same set of parameters as in subsection ~\ref{subsec:uhf-res} -- except that all the calculations are done solely for the realistic nickel-neodymium scaling parameter $\gamma = 1$. The obtained phase diagram can be seen in Fig.~\ref{fig:UHF-BRi}.

\begin{figure*}[t!]
\includegraphics[width=0.82\columnwidth,keepaspectratio]{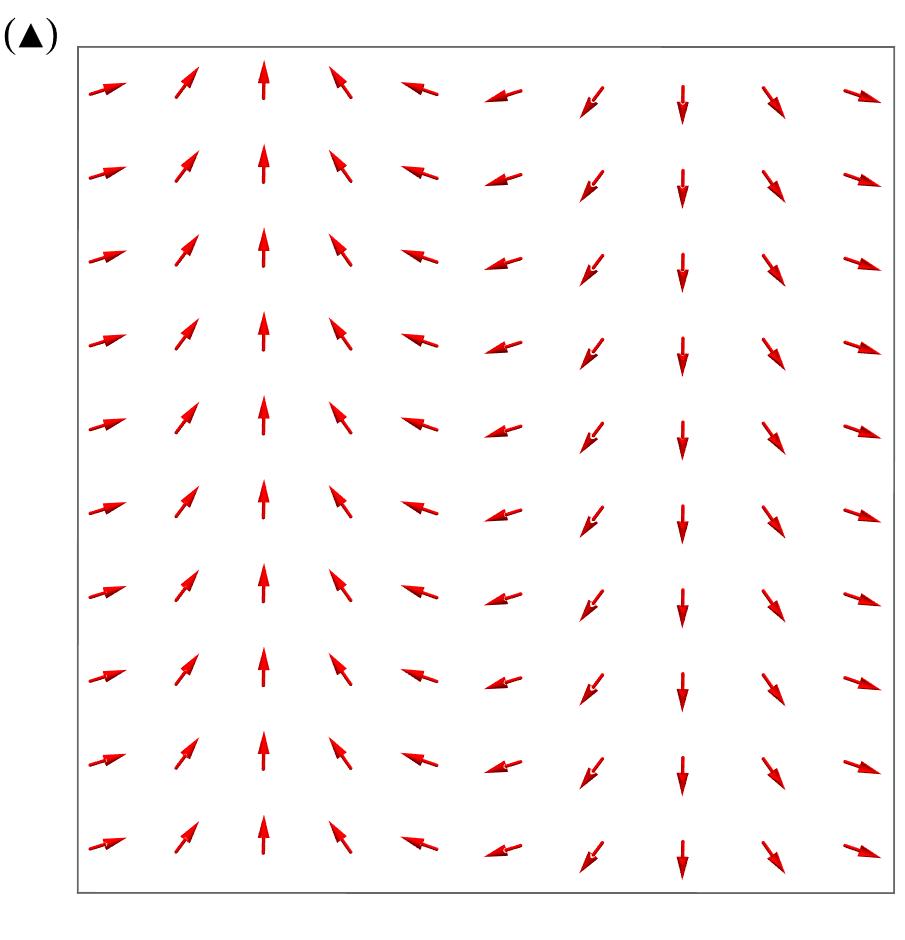}%
    \hskip 3mm
    \includegraphics[width=0.9\columnwidth,keepaspectratio]{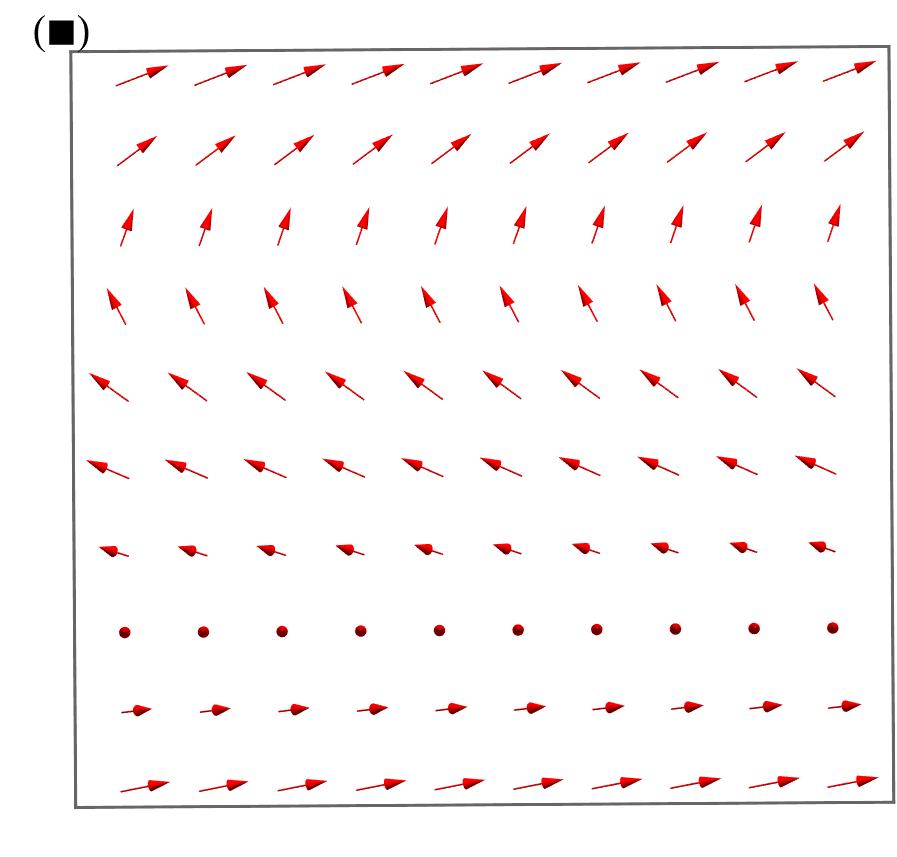}
    \vskip 3mm
    \includegraphics[width=0.87\columnwidth]{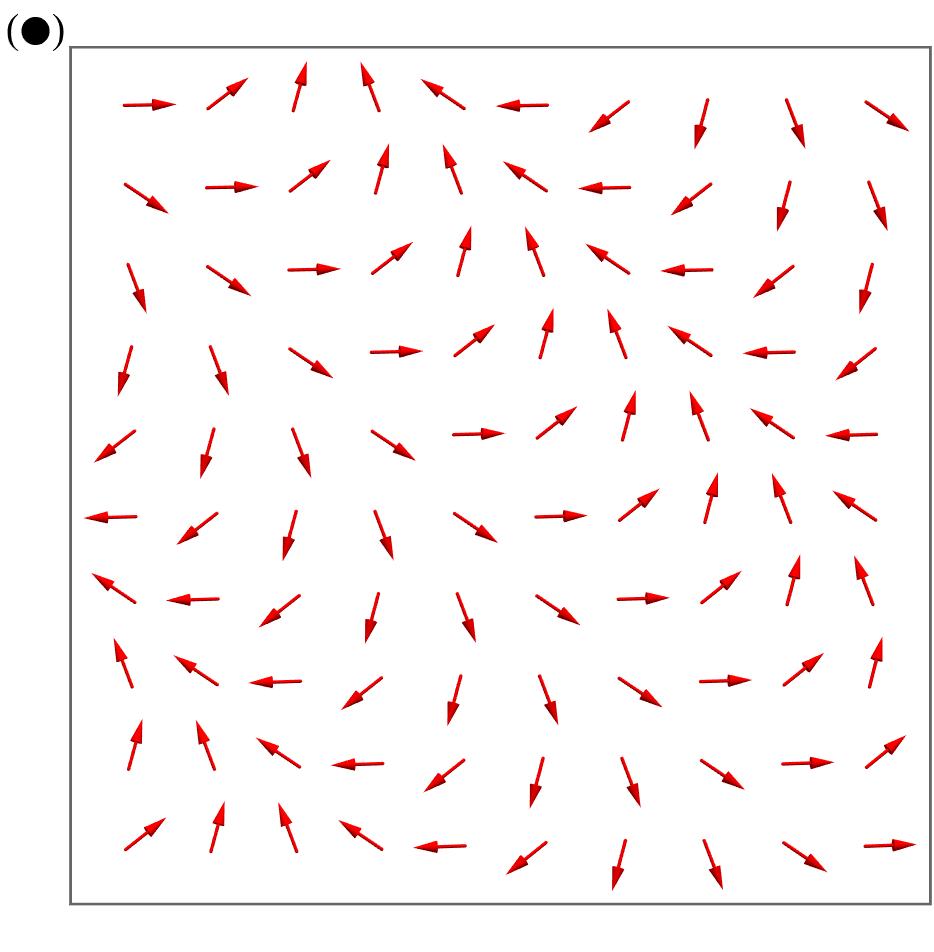}%
    \hskip 3mm
    \includegraphics[width=0.87\columnwidth]{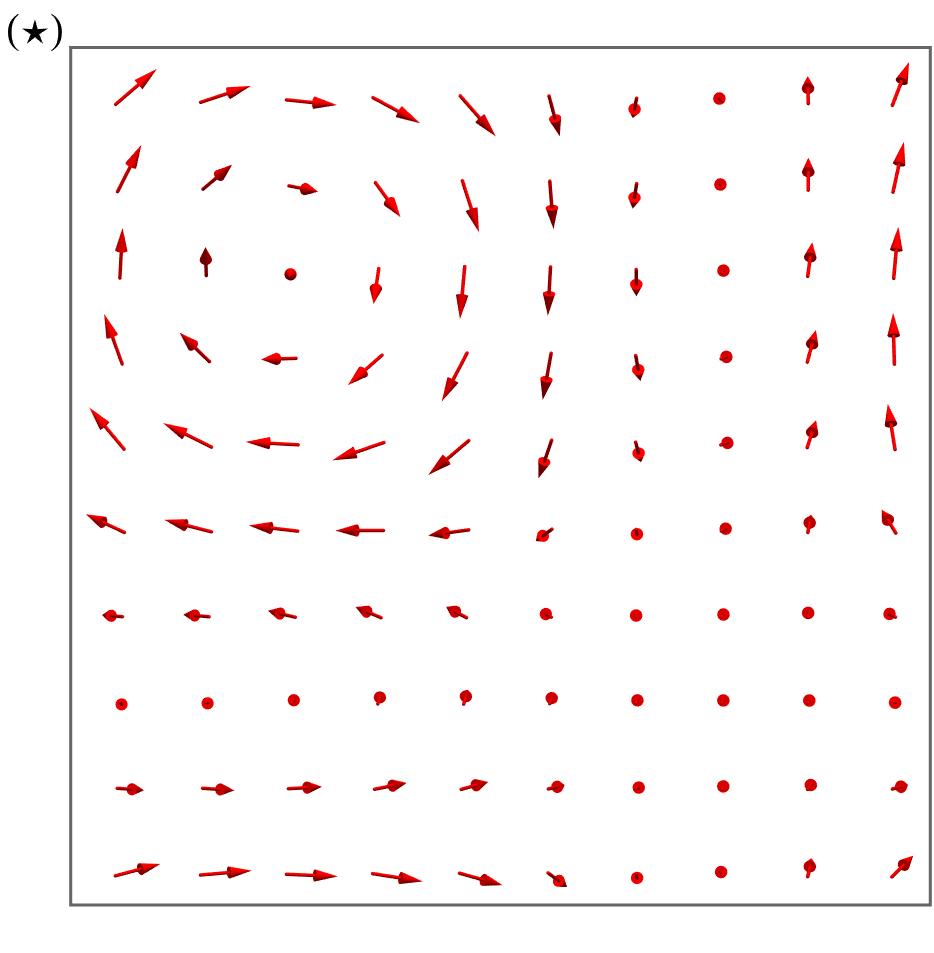}
    {
    \caption{Spin configurations 
    of the non-collinear orders,
    stable in the unrestricted 
    Hartree-Fock calculations 
    with the modified Broyden mixing
    of Sec.~\ref{subsec:uhf-broy}: 
    \textbf{($\blacktriangle$):} Spin configuration for the AFM axial stripe  order stable for charge transfer energy $\epsilon = 0.5$ eV and total filling $n = 1.15$. \textbf{($\blacksquare$):} Spin configuration for the FM axial stripe stable for $\epsilon = 0.5$ eV and $n = 1.15$. \textbf{($\bullet$):} Diagonal stipe spin configuration stable for $\epsilon = 1.1$ eV and $n = 1.4$. \textbf{($\star$):} Vortex spin configuration stable for for $\epsilon = 1.1$ eV and $n = 1.15$.
    }
    \label{fig:spins_vis}
    }
\end{figure*}

The results show that the AFM region is essentially unchanged relative to the results of the previous section, {\it cf}. Fig.~\ref{fig:UHF-BRi} {\it vs}. Fig.~\ref{fig:unrestricted-pd}. On the other hand, the phase space taken by the FM order diminishes. Instead various noncollinear phases set in. Interestingly, the `standard' UHF approach from the previous section only supports collinear phases.

The system may end up in various non-collinear orders, with stripes being more preferable for $\epsilon \leq 1.4$. This trend indicates that the small charge-transfer energy favors stripe order.
This is because lower charge transfer energy leads to finite self-doping on Ni$_{x^2-y^2}$ -- which enables the onset of stripes.
In fact, without the self-doping effect, the stripes would not occur -- this is visible from the phase diagram which shows that stripes occur solely within the FM phase, i.e. the phase that arises due to
self-doping (see next section for details).

Interestingly, the diagonal stripes (green in Fig.~\ref{fig:UHF-BRi}) and the axial stripes (magenta/red in Fig.~\ref{fig:UHF-BRi}) occur in roughly equal proportions, showing no strong orientation preference.
Fig.~\ref{fig:spins_vis} visualizes the three stable stripe and one incommensurate texture.

\section{Discussion} 
\label{sec:discussion}

\subsection{Ferromagnetic and antiferromagnetic phases} \label{subsec:fm-afm}

\begin{figure*}[t!]
    \begin{center}
        \huge \bf \hspace{40mm} FM \hfill AFM \hskip 34mm ~
    \end{center}
    \includegraphics[width=0.95\columnwidth]{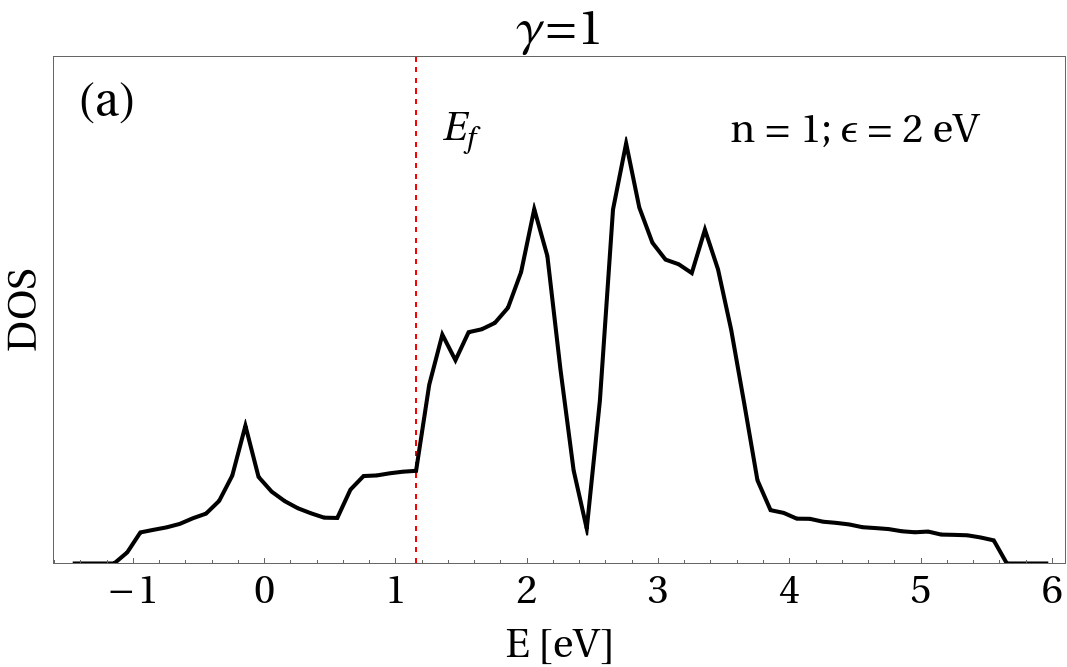}%
    \hskip 3mm
    \includegraphics[width=0.95\columnwidth]{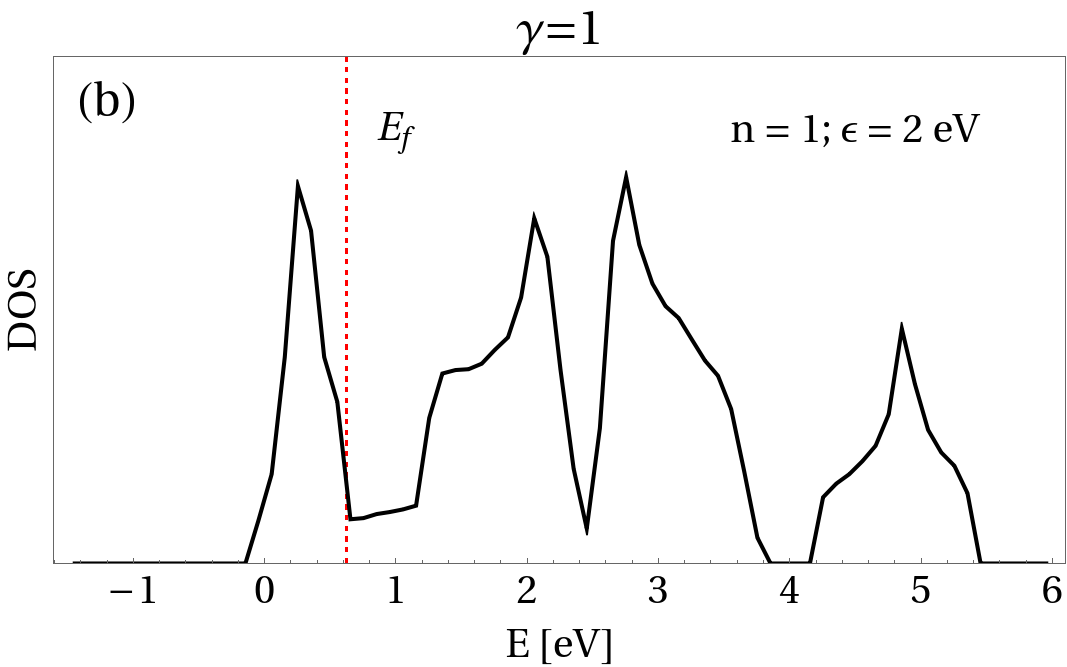}
    \vskip 3mm
    \includegraphics[width=0.95\columnwidth]{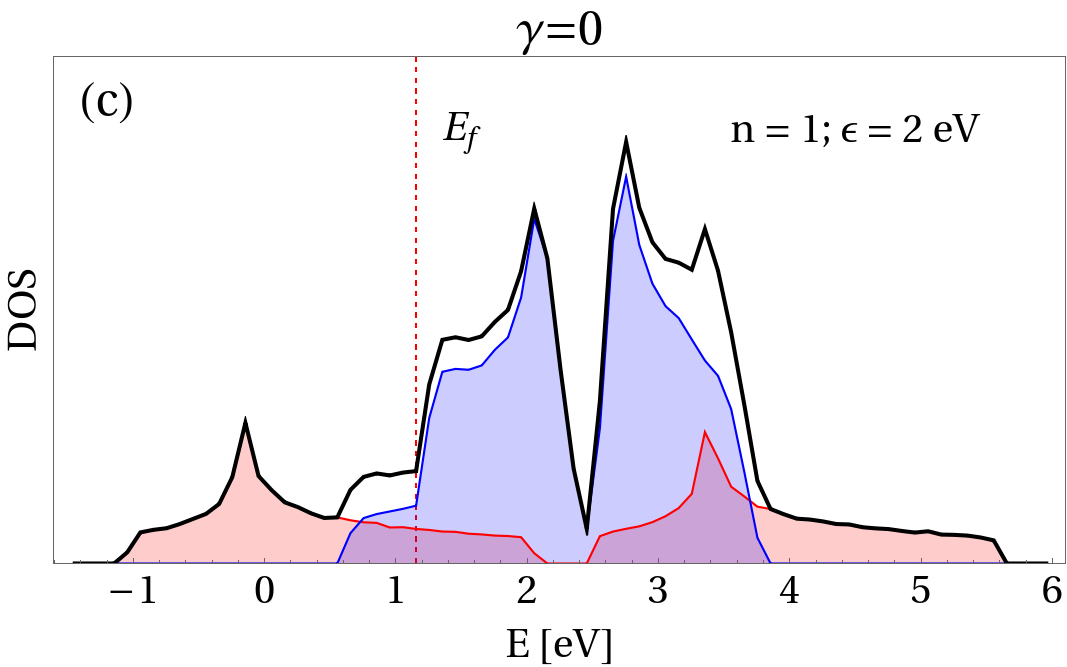}%
    \hskip 3mm
    \includegraphics[width=0.95\columnwidth]{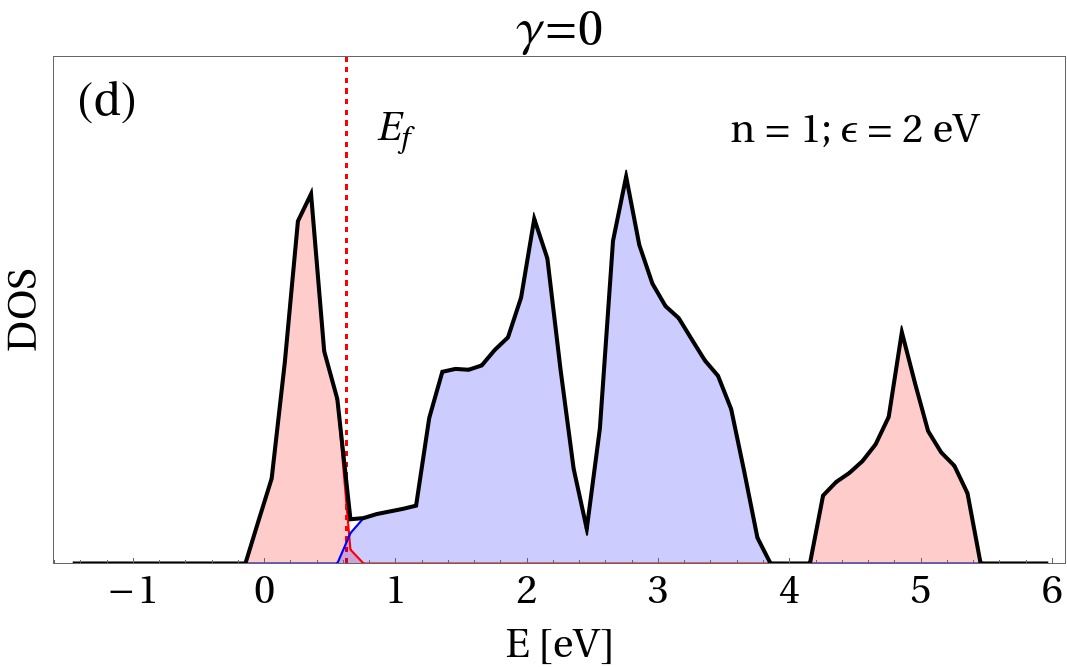}
    \vskip 3mm
    ~ \hskip 5mm \includegraphics[width=0.8\columnwidth]{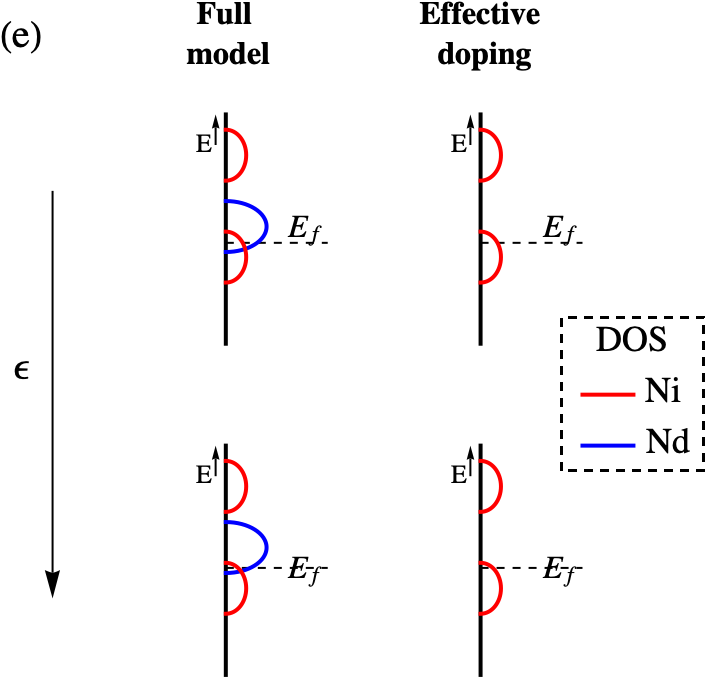}%
    \hskip 25mm
    \includegraphics[width=0.8\columnwidth]{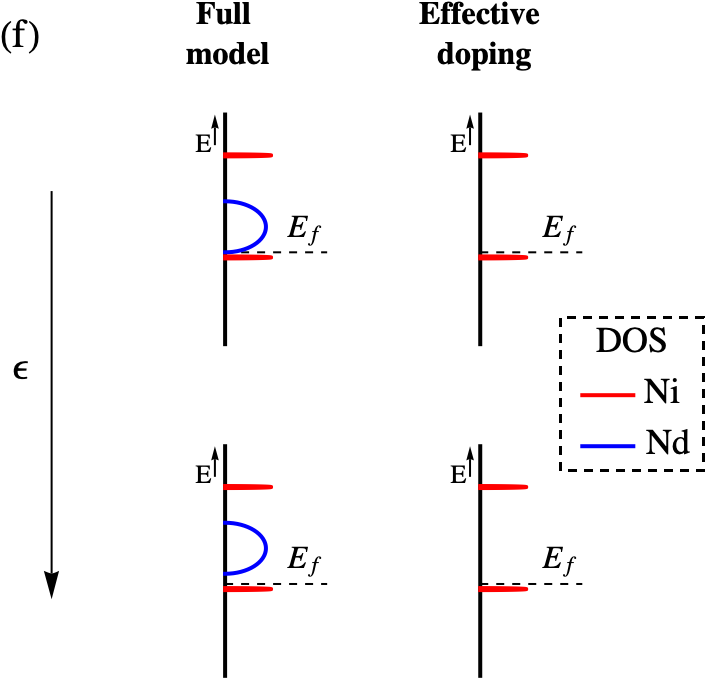}
    \caption{
    \textbf{Panels (a) and (b):} The density of states obtained by solving the Hamiltonian \eqref{eq:rhf-1-particle}-\eqref{eq:constant-rhf} in the restricted Hartree-Fock approximation, separately for the FM (a) and the AFM (b) case. The parameter values used: filling $n = 1$, charge transfer energy $\epsilon = 2$ eV and nickel-neodymium coupling $\gamma=1$.
    \textbf{Panels (c) and (d):} The same as (a)-(b), however here $\gamma = 0$ (the inter-layer coupling vanishes). The red and blue shaded areas represent the nickel and the neodymium DOS respectively. They sum to give the black line.
    \textbf{Panels (e) and (f):} A schematic picture showing how effective doping of the single-band Hubbard model describing the uncoupled nickel plane can be extracted from the full nickel-neodymium model in the FM case (e) and the AFM case (f).
    }
    \label{fig:stoner}
\end{figure*}
Let us begin by exploring the {\it origin} of the stability of the two dominant phases: the FM phase and the AF phase. 
Interestingly, in the RHF calculation we found that both the FM and the AFM solutions are stable in the {\it entire} $(\epsilon, n)$ parameter range~\eqref{eq:parameter-range-rhf} and just the relative energy decides which one is the ground state. This leads to a convenient situation, for we can compare the origin of these two ordered phases by looking at the solutions to the Hartree-Fock scheme for the same Hamiltonian parameters for  both phases.

The next crucial observation is that for both phases the nickel-neodymium hybridisation is qualitatively not important. Therefore, the nickel and neodymium bands can be considered as independent [see Fig.~\ref{fig:stoner} (a) vs (c) and (b) vs (d)].

The stability of the FM phase is due to the Stoner mechanism \cite{Stoner1939, Wysokinski2019}. The Coulomb interaction splits the nickel bands into an upper and a lower band, distinguished by the direction of spin [Fig.~\ref{fig:stoner}, panels (a), (c) and (e)]. The lower band's energy shift is equal to $U\rho_{0\downarrow}$, while the upper band is shifted by $U\rho_{0\uparrow}$ [see Eq.~\eqref{eq:rhf}; we have assumed that the spin-up direction is favoured by the Stoner mechanism]. Since (i) $U=4.5$ eV, which is comparable with the bandwidth of the neodymium bands, and (ii) the nickel on-site energy is lower than that of neodymium ($\epsilon > 0$), the neodymium density of states (DOS) is localized between the DOS of the two nickel Stoner bands [Fig. \ref{fig:stoner}(c, e)]. It acts as a reservoir for extra charge (self-doping), leaving the upper nickel Stoner band unoccupied for a wide range of fillings. This corresponds to perfect spin polarisation.

Just as in the FM case, the stability of the AFM phase is due to relatively large Hubbard $U$. In this case, the on-site Coulomb interaction also splits the nickel bands into an upper and a lower band [Fig.~\ref{fig:stoner}(b, d, f)], opening a substantial gap, equal to $2U\rho_1$. The neodymium DOS remains in that gap [Fig.~\ref{fig:stoner}(d, f)]. However, for the same realistic parameter values as in the FM case ($\epsilon=2$ eV, $n=1$), in the AFM case there is almost no overlap between the lower nickel band DOS and the neodymium DOS. Consequently, for $n \gtrsim 1$ all charge occupies the lower nickel band, leading to 
half-filling on nickel and stabilizing the AFM solution.

In order to get a feeling for the phase diagram presented in Fig.~\ref{fig:restricted-pd}, let us consider the effect of varying the free parameters in our model away from their physical values for an undoped system, namely away from $\epsilon = 2$ eV and $n=1$.

\textit{Varying $\epsilon$, $n=1$.} Changing the charge transfer energy $\epsilon$ [the vertical axis of Fig.~\ref{fig:restricted-pd}] amounts to shifting the neodymium DOS upwards or downwards in energy [see Fig.~\ref{fig:stoner}, panels (e) and (f)]. In the FM case, for $\epsilon <$ 2eV, i.e. lower than the realistic value of $\epsilon=2$ eV, the overlap between the lower nickel Stoner band DOS and the neodymium DOS is increased;  on the other hand, for $\epsilon > 2$ eV the overlap is decreased. This decrease in the overlap for increasing $\epsilon$ leads to the lower nickel band eventually becoming fully occupied and the FM state becoming energetically less favorable (see Fig.~\ref{fig:restricted-pd}) and ultimately unstable for large enough $\epsilon$ (unshown).

In the AFM case, for the physical filling $n=1$ and for $\epsilon \gtrsim 2$ eV, there is no overlap between the lower nickel band DOS and the neodymium DOS [see Fig.~\ref{fig:stoner}, panels (d) and (f)]. All charge remains on the lower nickel band. Decreasing the nickel-neodymium charge transfer energy below $\epsilon = 2$ eV introduces an overlap between the lower nickel band DOS and the neodymium DOS, leading to the escape of charge onto neodymium. This constitutes hole doping of the nickel plane and leads to the AFM phase being less energetically favorable than the FM phase (see Fig.~\ref{fig:restricted-pd}) with decreasing $\epsilon$ and ultimately to the destabilisation of the AFM ground state for small enough $\epsilon$ (unshown). This is why in the phase diagram in Fig.~\ref{fig:restricted-pd} the AFM phase is stable for higher values of $\epsilon$. 

\textit{Varying $n$, $\epsilon=2$.} Changing the filling $n$ [horizontal axis in Fig.~\ref{fig:restricted-pd}] will have the following effect: (i) If the lower nickel band is already full, the extra charge will be absorbed by the neodymium DOS located between the nickel bands, thus leading to no change in the filling of the nickel plane [see Fig.~\ref{fig:stoner}, panels (d) and (f)] and no change in magnetism (AFM state stable as long as the lower nickel band is full); (ii) If the lower nickel band is not fully occupied, increasing $n$ will eventually fill it [see Fig.~\ref{fig:stoner}, panels (c) and (e)], supporting the AFM phase over the FM phase. This is why in the phase diagram in Fig.~\ref{fig:restricted-pd} the phase boundary is `tilted' with respect to the horizontal axis.

\subsection{The phase boundary} \label{subsec:phase-boundary}

In the UHF results non trivial magnetic orders are found at the FM/AFM phase boundary. These represent the coexistence of FM and AFM phases, as the magnetic patterns come in the shape of FM islands immersed in an AFM sea, as well as more regular, striped patterns [see insets in Fig.~\ref{fig:unrestricted-pd} (a)-(c)]. 
However, the studies we present here were done on a cluster of size 10x10 with PBC. For larger clusters the coexistence of phases seen in Fig.~\ref{fig:unrestricted-pd} might not survive. Indeed, by performing calculations for larger clusters for such points $(\epsilon, n)$ that yield exotic orders in the phase diagram shown in Fig.~\ref{fig:unrestricted-pd}, we have found that for large enough clusters either FM or AFM orders stabilize in the phase boundary region, suggesting that the coexistence of phases is a finite size effect on the Hartree-Fock level. 

We note already here that one may expect that going beyond the Hartree-Fock level, this near-degeneracy of the competing phases leads to a spin-liquid like solution. Intriguingly, the physical values of the two free parameters in our model, the nickel-neodymium charge transfer energy $\epsilon = 2$ eV and the total filling $n = 1$~\cite{Wu2020}, lie right in the region where this complex magnetic picture emerges in the mean-field calculations. Thus, we further comment on this important issue in the end of the next subsection and in the Conclusions.

\subsection{Validity of the results}

In principle, different variants of mean-field methods have been quite successfully used to study correlated systems~\cite{Auerbach1994, Sachdev2023}. Nevertheless, a lot of care has to be taken when interpreting the results.

The {\it first} point is that our results show that the neodymium orbitals can essentially be regarded as charge reservoir for doping the single-band Hubbard model, i.e. the nickel-neodymium hybridisation effect is largely irrelevant. While this stays in agreement with the {\it ab-initio} results for NdNiO$_2$ of \cite{Kitatani2020} and it seems plausible that this result
indeed survives beyond mean-field studies, this crucial result naturally requires further studies to be confirmed.

The {\it second} point concerns solutions of the 2D doped single-band Hubbard model. 
Here our results are in a general agreement with various
mean-field studies which point to the stability
of both the AFM~\cite{Laughlin2014, Scholle2023} and 
the FM phases~\cite{Bach2006} in the doped Hubbard model.
Note that in our case the onset of the FM phase is due to the rather substantial longer-range hopping. However, the crucial question is to what extent the FM and the AFM phase survive beyond the mean-field approximation:

While it is quite well-established that AFM may exist in the (lighty-)doped Hubbard models ~\cite{Dagotto1992, Koepsell2021}, the FM solution is more problematic. In general, the search for the mechanism behind itinerant ferromagnetism continues, and it is safe to say that the phenomenon is still not understood~\cite{Vollhardt1999,Santiago2017}. The stability of the FM solution in correlated models is usually exaggerated by mean-field calculations, that support the Stoner mechanism~\cite{Stoner1939, Bach2006, Wysokinski2019}. In fact, it has been shown that using a more correlated treatment yields no FM solution in the square-lattice, single-band Hubbard model~\cite{Hirsch1985}.
Nevertheless, tendency to FM in the Hubbard model with longer-range hopping was observed~\cite{Farkaovsk2002}. Last but not least, the Nagaoka theorem~\cite{Nagaoka1966} also guarantees the onset of FM in the doped Hubbard model -- albeit rather in an extremely correlated and lightly doped limit that is of no relevance here.

Nevertheless, the {\it third} point we want to make here is that, even if the stability of the FM phase is a purely mean-field effect, still the calculated degeneracy of the FM and AFM phases found at the phase boundary has important consequences.
Namely, it shows that at, and naturally also near, the phase boundary distinct terms in the Hamiltonian strongly compete. Typically such frustrated interactions lead to the enhanced role of quantum fluctuations and suggest a possible collapse of the LRO order close to the phase boundary calculated in these mean-field studies.

\section{Conclusions}

Our calculations show that the self-doping effect in the infinite-layer nickelate NdNiO$_2$ can be explained by the neodymium states in the gap between the upper and lower nickel bands. These bands are split by Coulomb interaction and the size of the gap is proportional to $U$. The neodymium states to a large extent act as charge reservoir for doping the single-band Hubbard model. This stays in agreement with the results
of Ref.~\cite{Kitatani2020}. 

We found that physically realistic values of the two free parameters in our model, the nickel-neodymium charge-transfer energy $\epsilon =2$ eV and the total filling $n=1$, put NdNiO$_2$ right outside the AFM region on the phase diagram [see Fig.~\ref{fig:unrestricted-pd}]. For a finite cluster, this region corresponds to a coexistence of AFM and FM phases, suggesting that the ground state is almost degenerate. This leads to the suppression of LRO magnetism and enhanced
quantum fluctuations. However, the proximity to the AFM LRO might still be detectable, for instance in the presence of para-magnons~\cite{Tacon2011,  Dean2013, Jia2014, Mitrano2024}. Indeed, that is what is seen in experiment~\cite{Lu2021,Fowlie2022}. 

Moreover, we have demonstrated that, because of this neodymium `charge reservoir' absorbing extra electrons, the nickel plane is half filled for a wide range of electron dopings. This leads to the `re-entrance' to the AFM upon electron-doping
NdNiO$_2$ [see Fig.~\ref{fig:unrestricted-pd}]. It is clear that in the case of \textit{hole} doping an analogous situation would not occur. Any holes added to NdNiO$_2$ would be absorbed by the Ni-O plane, and the Ni-O subsystem would immediately deviate from half-filling, destabilising the AFM solution.

It is also interesting to note that the more complex Hartree-Fock calculations (using the Broyden mixing method) may also support the
onset of stripes. This shows that stripes can become stable already on a static mean-field level and the more advanced treatment
of correlations on the DMFT level is not needed to observe stripes in these systems~\cite{Chen2023}.
The onset of stripes occurs 
for relatively low charge-transfer energies and for particular fillings. While in general stripes can always be expected in the doped Hubbard-like system, it is interesting to note that here they actually arise due to the self-doping effect.
Note, however, that for a realistic case of the charge transfer energy $\epsilon = 2$ eV, i.e. being not too low,
the self-doping effect is small and according to the mean-field calculations of this work
the stripes cannot become stable neither for pristine nor for electron-doped NdNiO$_2$.

As a final remark, we emphasize that these results were obtained using a model with realistic kinetic energy~\cite{Wu2020} and a value for the Hubbard on-site interaction $U$~\cite{Rosa2024}. Furthermore, this thorough study of the unrestricted HF of the extended Hubbard model for nickelates complements the recent unrestricted HF study of the 2D Hubbard model for cuprates of Ref. \cite{Scholle2023}.

\section*{Acknowledgements}

We thank Karsten Held, Marcin Wysokiński and Andrzej M. Ole\'s for 
stimulating discussions.
We kindly acknowledge support from the
National Science Centre (NCN, Poland) under Project
No. 2021/43/B/ST3/02166
{ and 2024/55/B/ST3/03144}. The work is 
supported by the Foundation for Polish Science through the 
IRA Programme co-financed by EU within SG OP Programme.

For the purpose of Open Access, the authors have applied a CC-BY 
public copyright licence to any Author Accepted Manuscript (AAM) 
version arising from this submission.

The data and scripts to reproduce figures presented in this manuscript are available at Ref.~\cite{zenodo}.

\begin{appendix}

\section{The hopping matrices $k_{i,j}$} \label{sec:appendix1}

Below we give values of the hopping matrices $k_{ij}$
that perfectly reproduce the tight-binding model of Ref.~\cite{Wu2020} truncated to a single nickel and neodymium layer. Note the properties $k_{i,j}=k_{-i,-j}^T$ and $k_{i,j}=g\,k_{j,i}\,g$ with $g$ being a diagonal matrix with the entries $\{1,1,-1\}$
and that all entries are given in eV. We have:
\begin{align} \label{eq:matrices}
k_{0,0}&=\begin{pmatrix}2.3692 & 0 & 0\\
0 & 2.4463 & 0\\
0 & 0 & 2.3692 - \epsilon
\end{pmatrix},\\
k_{1,0}&=\begin{pmatrix}-0.387 & 0 & 0\\
0 & 0.3202 & 0\\
0.0219 \gamma & -0.0139 \gamma & -0.3761
\end{pmatrix},\nonumber\\
k_{2,0}&=\begin{pmatrix}-0.034 & 0 & 0.0219 \gamma\\
0 & 0.0367 & 0.0139 \gamma\\
0 & 0 & -0.0414
\end{pmatrix},\nonumber\\
k_{3,0}&=\begin{pmatrix}0 & 0 & 0\\
0 & 0.012 & 0\\
0 & 0 & 0
\end{pmatrix},\nonumber\\
k_{1,1}&=\begin{pmatrix}0 & 0.0798 & 0\\
0.0798 & -0.0467 & 0\\
0 & 0 & 0.0844
\end{pmatrix},\nonumber\\
k_{1,2}&=\begin{pmatrix}0 & 0 & -0.0219 \gamma\\
0 & -0.0198 & 0.0139 \gamma\\
0 & 0 & -0.0043
\end{pmatrix},\nonumber\\
k_{2,2}&=k_{-2,2}=\begin{pmatrix}0 & 0 & 0\\
0 & 0 & 0\\
0 & 0 & 0.003
\end{pmatrix},\nonumber\\
k_{-1,1}&=\begin{pmatrix}0 & -0.0798 & 0.0219 \gamma\\
-0.0798 & -0.0467 & 0.0139 \gamma\\
-0.0219 \gamma & -0.0139 \gamma & 0.0844
\end{pmatrix},\nonumber\\
k_{-1,2}&=\begin{pmatrix}0 & 0 & 0\\
0 & -0.0198 & 0\\
0 & 0 & -0.0043
\end{pmatrix}.\nonumber
\end{align}
Above, $\gamma$ is the scaling of the inter-layer, nickel-neodymium hopping and $\epsilon$ is the nickel-neodymium charge transfer energy.

\end{appendix}

\end{document}